\documentclass[english]{article}
\usepackage[T1]{fontenc}
\usepackage[latin9]{inputenc}
\usepackage{geometry}
\usepackage{amsmath}
\usepackage{graphicx}
\usepackage{esint}

\makeatletter
\@ifundefined{date}{}{\date{}}

\usepackage{float}
\usepackage{textcomp}

\usepackage{babel}

\IfFileExists{lmodern.sty}{\usepackage{lmodern}}{}

\usepackage{babel}

\@ifundefined{showcaptionsetup}{}{%
 \PassOptionsToPackage{caption=false}{subfig}}
\usepackage{subfig}
\makeatother

\usepackage{babel}
\begin{document}

\title{Diphoton decay for a $750$ GeV scalar boson in an $U(1)_{X}$ model}

\author{R. Martinez\thanks{remartinezm@unal.edu.co}, F. Ochoa\thanks{faochoap@unal.edu.co},
C.F. Sierra\thanks{cfsierraf@unal.edu.co} }

\maketitle
\begin{center}
\textit{Departamento de Física, Universidad Nacional de Colombia,
Ciudad Universitaria, K. 45 No. 26-85, Bogotá D.C., Colombia} 
\par\end{center}
\begin{abstract}
In the context of a nonuniversal and anomaly free $U(1)_{X}$ extension
of the standard model, we examine the decay of a $750$GeV scalar
singlet state, $\xi_{\chi}$, as a possible explanation of the observed
diphoton excess announced by the ATLAS and CMS collaborations at CERN-LHC
collider. One-loop decay to photons is allowed through three heavy
singlet quarks and one charged Higgs boson into the loop. We obtain,
for different width approximations and for masses of the exotic singlet
quarks in the region $[900,\,3000]$ GeV, a production cross section
$\sigma(pp\to\xi_{\chi}\to\gamma\gamma)$ compatible with ATLAS and
CMS collaborations data. We also include another scalar singlet, $\sigma$,
as a dark matter candidate that may couple with the 750 GeV scalar
at tree level with production cross sections in agreement with ATLAS
and CMS. 
\end{abstract}

\section{Introduction}

\label{intro} Although the Standard Model (SM) \cite{SM} is the
simplest model that successfully explains most of the phenomena and
experimental observations in particle physics, there are still some
unexplained and unanswered fundamental questions which many theorists
associate with an underlying theory beyond the SM. The most recent
experimental discrepancy is the 3$\sigma$ excess in the diphoton
channel at $750$ GeV announced by the ATLAS and CMS collaborations
\cite{CMS750,ATLAS750} which has been the subject of many interpretations
in the literature from different extensions of the standard model
(SM) \cite{key-3,key-4,key-5,key-6,key-7,key-8,key-9,key-10,key-11}.
Although this observation requires further analyses and more experimental
data, it is interesting to explore the theoretical and phenomenological
consequences to have a new resonance with this mass. In particular,
a scalar candidate is supported by many theoretical models, as for
example, heavier Higgs bosons from scalar extensions of the SM, as
recently considered in \cite{chiara}, models with heavy axion candidates
as shown in \cite{Pilaftsis} and with pseudo-Nambu Goldstone bosons
as in \cite{1512.04850}. Authors in \cite{ellis} and \cite{florian}
have studied general cases for different possible models.

In particular, nonuniversal $U(1)'$ symmetry models have many well-established
motivations. First, since family representations are nonuniversal,
they may provide hints for solving the SM flavor puzzle \cite{flavorphysics}.
Secondly, these models contains an extra $Z'$ neutral gauge boson
with many interesting phenomenological consequences at low and high
energies \cite{zprime-review}. In some models, an extended fermion
spectrum is necessary in order to obtain an anomaly-free theory, providing
a natural scenario for extra charged leptons and/or heavy quarks.
Also, the new symmetry requires an extended scalar sector in order
to \textit{i.)} generate the breaking of this symmetry and \textit{ii.)}
obtain heavy masses for the new $Z'$ gauge boson and the extra fermion
content. Another consequence of an extended Higgs sector is that they
may produce deviations of the Higgs self-coupling, which could provide
an interesting test for the SM Higgs boson from future measurements
at the LHC collider \cite{trilineal}.

In this paper, we evaluate the process of a 750 GeV scalar particle
decaying into two photons in the context of the nonuniversal $U(1)_{X}$
extension introduced in Refs. \cite{modelo3,modelo2,modelo1}, which
gives us a natural scenario with one-loop contributions from heavy
quarks and charged Higgs bosons. In section 2 we present the particle
content of the model as well as the Higgs potential and the Yukawa
Lagrangian. In section 3, we analyze the diphoton decay by using three
approximations for the decay width. First, we assume that the total
decay of the scalar candidate come only from one loop decay contributions.
Second, we take the total width as $\Gamma=45$GeV, reported by the
ATLAS Collaboration. Finally, we consider the decay into a scalar
dark matter candidate, $\sigma$.

\section{Description of the model}

\label{sec:1} We consider the abelian extension $G_{sm}\times U(1)_{X}$,
where $G_{sm}=(SU(3)_{c},SU(2)_{L},U(1)_{Y})$ is the ordinary SM
gauge symmetries, while $U(1)_{X}$ is an extra symmetry that assign
a new charge $X$ to the particle content, as shown in tables \ref{tab:SM-espectro}
and \ref{tab:exotic-espectro}. Some general properties of the model
are:

\begin{table}
\caption{{\small{}Ordinary SM particle content, with $i=$1,2,3\label{tab:SM-espectro}}}
\label{tab:1}{\small{} 
\[
\begin{tabular}{lll}
\hline  \noalign{\smallskip} \ensuremath{Spectrum}  &  \ensuremath{G_{sm}}  &  \ensuremath{U(1)_{X}} \\
 \\
\hline  \noalign{\smallskip}\noalign{\smallskip} \hspace{-0.3cm}\ensuremath{\ \begin{tabular}{l}
 \ensuremath{q_{L}^{i}=\left(\begin{array}{c}
U^{i}\\
D^{i}
\end{array}\right)_{L}} \end{tabular}\ }  &  \ensuremath{(3,2,1/3)}  &  \hspace{-0.3cm}\ensuremath{\ \begin{tabular}{c}
 \ensuremath{1/3} for \ensuremath{i=1} \\
 \ensuremath{0} for \ensuremath{i=2,3} 
\end{tabular}\ }\\
  &   &  \\
 \ensuremath{U_{R}^{i}}  &  \ensuremath{(3,1,4/3)}  &  2/3\\
  &   &  \\
 \ensuremath{D_{R}^{i}}  &  \ensuremath{(3,1,-2/3)}  &  \ensuremath{-1/3}\\
  &   &  \\
 \hspace{-0.3cm}\ensuremath{\ \begin{tabular}{c}
 \ensuremath{\ell_{L}^{i}=\left(\begin{array}{c}
\nu^{i}\\
e^{i}
\end{array}\right)_{L}}\end{tabular}\ }  &  \ensuremath{(1,2,-1)}  &  \ensuremath{-1/3}\\
  &   &  \\
 \ensuremath{e_{R}^{i}}  &  \ensuremath{(1,1,-2)}  &  \begin{tabular}{c}
 \ensuremath{-1}\end{tabular}\\
  &   &  \\
 \hspace{-0.3cm} \ensuremath{\ \begin{tabular}{c}
 \ensuremath{\phi_{1}=\left(\begin{array}{c}
\phi_{1}^{+}\\
\frac{\upsilon_{1}+\xi_{1}+i\zeta_{1}}{\sqrt{2}}
\end{array}\right)}\end{tabular}\ }  &  \ensuremath{(1,2,1)}  &  \ensuremath{2/3} 
\\\hline \end{tabular}
\]
}
\end{table}

\begin{itemize}
\item The equations that cancel the chiral anomalies are obtained in \cite{modelo3}.
These equations leads us to a set of non-trivial solutions for $U(1)_{X}$
that requires a structure of three families. First, the left-handed
leptons $\ell_{L}^{i}$ are universal of family, with charge $X_{\ell}=-1/3$.
Second, the left-handed quarks $q_{L}^{i}$ have nonuniversal charges:
family with $i=1$ has $X_{1}=1/3$, while $X_{2,3}=0$ for $i=2,3$.
In addition, the cancellation of anomalies require the existence of
an extended fermion sector. A simple possibility in the quark sector
is by introducing quasi-chiral singlets ($T$ and $J^{n}$, where
$n=1,2$), i.e. singlets that are chiral under $U(1)_{X}$ and vector-like
under the SM. 
\item An extra neutral gauge boson, $Z'_{\mu}$, is required to make the
$U(1)_{X}$ transformation a local symmetry. 
\end{itemize}
\begin{table}
\caption{{\small{}Extra non-SM particle content, with $n=$1,2\label{tab:exotic-espectro}}}
\label{tab:2}{\small{}
\[
\begin{tabular}{lll}
\hline  \noalign{\smallskip} \ensuremath{Spectrum}  &  \ensuremath{G_{sm}}  &  \ensuremath{U(1)_{X}} \\
\hline  \noalign{\smallskip}\noalign{\smallskip} \ensuremath{T_{L}}  &  \ensuremath{(3,1,4/3)}  &  1/3\\
  &   &  \\
 \ensuremath{T_{R}}  &  \ensuremath{(3,1,4/3)}  &  2/3\\
  &   &  \\
 \ensuremath{J_{L}^{n}}  &  \ensuremath{(3,1,-2/3)}  &  0\\
  &   &  \\
 \ensuremath{J_{R}^{n}}  &  \ensuremath{(3,1,-2/3)}  &  -1/3\\
  &   &  \\
 \hspace{-0.3cm} \ensuremath{\ \begin{tabular}{c}
 \ensuremath{\phi_{2}=\left(\begin{array}{c}
\phi_{2}^{+}\\
\frac{1}{\sqrt{2}}(\upsilon_{2}+\xi_{2}+i\zeta_{2})
\end{array}\right)}\end{tabular}\ }  &  \ensuremath{(1,2,1)}  &  \ensuremath{1/3} \\
  &   &  \\
 \hspace{-0.3cm} \begin{tabular}{c}
 \ensuremath{\chi=\frac{1}{\sqrt{2}}(\upsilon_{\chi}+\xi_{\chi}+i\zeta_{\chi})}\end{tabular}  &  \ensuremath{(1,1,0)}  &  \ensuremath{-1/3} \\
  &   &  \\
 \hspace{-0.3cm} \begin{tabular}{c}
 \ensuremath{\sigma}\end{tabular}  &  \ensuremath{(1,1,0)}  &  \ensuremath{-1/3} \\
  &   &  \\
 Z\ensuremath{'_{\mu}}  &  (1,1,0)  &  0 \\
  &   &  \\
 \hspace{-0.3cm} \begin{tabular}{c}
 \ensuremath{(\nu_{R}^{i})^{c}}\\
 \\
 
\end{tabular}  &  \ensuremath{(1,1,0)}  &  \ensuremath{-1/3}\\
  &   &  \\
 \hspace{-0.3cm} \begin{tabular}{c}
 \ensuremath{N_{R}^{i}}\end{tabular}  &  \ensuremath{(1,1,0)}  &  \ensuremath{0} 
\\\hline \end{tabular}
\]
}
\end{table}

\begin{itemize}
\item Due to the nonuniversal structure of the quark doublets, two scalar
doublets $\phi_{1}$ and $\phi_{2}$ identical under $G_{sm}$ but
with $U(1)_{X}$ charges $X_{\phi_{1}}=2/3$ and $X_{\phi_{2}}=1/3$,
respectively, are required in order to obtain massive fermions after
the spontaneous symmetry breaking, where the electroweak vacuum expectation
value (VEV) is $\upsilon=\sqrt{\upsilon_{1}^{2}+\upsilon_{2}^{2}}$. 
\item An extra scalar singlet $\chi$, with $U(1)_{X}$ charge $X=-1/3$
and VEV $\upsilon_{\chi}$ is required to produce the symmetry breaking
of the $U(1)_{X}$ symmetry. We assume that it happens at a large
scale $\upsilon_{\chi}>\upsilon$. The real component $\xi_{\chi}$
remains in the particle spectrum after the symmetry breaking, and
is our candidate to explain the $750$ GeV signal. The imaginary component
$\zeta_{\chi}$ is the would-be Goldstone boson that provides mass
to the extra neutral gauge boson $Z'$. 
\item Another scalar singlet, $\sigma$, is introduced, which is a scalar
dark matter (DM) candidate. In order to reproduce the observed DM
relic density, this particle must accomplish the following minima
conditions \cite{modelo2,modelo1}:

\begin{enumerate}
\item[(i)] Since $\sigma$ acquires a nontrivial charge $U(1)_{X}$, it must
be complex in order to be a massive candidate. 
\item[(ii)] To avoid odd powers terms in the scalar Lagrangian, which leads to
unstable DM, we impose the global continuous symmetry

\begin{eqnarray}
\sigma\rightarrow e^{i\theta}\sigma.\label{global-symm}
\end{eqnarray}

\item[(iii)] In spite of the above symmetry, the model still can generate odd
power terms via spontaneous symmetry breaking. To avoid this, $\sigma$
must not generate VEV during the lifetime of our Universe. 
\end{enumerate}
\item Extra discrete symmetries can be assumed in this model for scalar
and quarks fields in order to obtain hierarchical mass structures,
as shown in \cite{modelo3}. However, these types of symmetries do
not affect the Yukawa couplings that participates in the diphoton
signal. Thus, additional global symmetries will be irrelevant in our
calculations. 
\item It is desirable to obtain a realistic model compatible with the oscillation
of neutrinos. For this purpose, the model introduces new neutrinos,
$(\nu_{R}^{i})^{c}$ and $N_{R}^{i}$ with $i=1,2,3$ which may generate
seesaw neutrino masses. However, this sector will be irrelevant in
the present analysis. 
\end{itemize}

\subsection{Higgs potential}

\label{sec:2} 

As shown in \cite{modelo2}, the most general, renormalizable and
$G_{sm}\times U(1)_{X}$ invariant potential with the symmetry $\sigma\rightarrow e^{i\theta}\sigma$
is

\begin{eqnarray}
V & = & \mu_{1}^{2}\left|\phi_{1}\right|^{2}+\mu_{2}^{2}\left|\phi_{2}\right|^{2}+\mu_{3}^{2}\left|\chi\right|^{2}+\mu_{4}^{2}\left|\sigma\right|^{2}\nonumber \\
 & + & f_{2}\left(\phi_{2}^{\dagger}\phi_{1}\chi+h.c.\right)\nonumber \\
 & + & \lambda_{1}\left|\phi_{1}\right|^{4}+\lambda_{2}\left|\phi_{2}\right|^{4}+\lambda_{3}\left|\chi\right|^{4}+\lambda_{4}\left|\sigma\right|^{4}\nonumber \\
 & + & \left|\phi_{1}\right|^{2}\left[\lambda_{6}\left|\chi\right|^{2}+\lambda'_{6}\left|\sigma\right|^{2}\right]\nonumber \\
 & + & \left|\phi_{2}\right|^{2}\left[\lambda_{7}\left|\chi\right|^{2}+\lambda'_{7}\left|\sigma\right|^{2}\right]\nonumber \\
 & + & \lambda_{5}\left|\phi_{1}\right|^{2}\left|\phi_{2}\right|^{2}+\lambda'_{5}\left|\phi_{1}^{\dagger}\phi_{2}\right|^{2}+\lambda_{8}\left|\chi\right|^{2}\left|\sigma\right|^{2}.\label{higgs-pot-1}
\end{eqnarray}

\noindent When we apply the minimum conditions $\partial\langle V\rangle/\partial\upsilon_{i}=0$
for each scalar VEV $\upsilon_{i}=\upsilon_{1,2,\chi}$, following
\cite{modelo2} we obtain at dominant order

\begin{eqnarray}
\mu_{3}^{2} & \approx & -\lambda_{3}\upsilon_{\chi}^{2},
\end{eqnarray}
which will allow us to obtain the mass of the real component of the
scalar $\chi$. With the above conditions, we can obtain the squared
mass matrices, for neutral real, neutral imaginary and charged scalar
components. After diagonalization, we obtain three scalar mass eigenstates
$(h,H,H_{\xi})$ from the real mass matrix, one pseudoscalar boson
$A$ from the imaginary matrix and one charged scalar $H^{\pm}$ from
the charged matrix. As shown in equation (\ref{scalar-eigenvectors}),
the scalar Higgs boson $H_{\xi}$ is identified with the real component
$\xi_{\chi}$, which is our $750$ GeV candidate. There are also would-be
Goldstone bosons that are absorbed as longitudinal components of the
charged weak bosons $W^{\pm}$, and the two neutral gauge bosons $Z$
and $Z'$. In the end, we obtain the following mass eigenvectors:

\begin{eqnarray}
\begin{pmatrix}G^{\pm}\\
H^{\pm}
\end{pmatrix} & = & R_{\beta}\begin{pmatrix}\phi_{1}^{\pm}\\
\phi_{2}^{\pm}
\end{pmatrix},\ \ \ \begin{pmatrix}G\\
A
\end{pmatrix}=R_{\beta}\begin{pmatrix}\zeta_{1}\\
\zeta_{2}
\end{pmatrix},\nonumber \\
\begin{pmatrix}h\\
H
\end{pmatrix} & = & R_{\alpha}\begin{pmatrix}\xi_{1}\\
\xi_{2}
\end{pmatrix},\ \ \ \ H_{\chi}\approx\xi_{\chi},\ \ \ \ G_{\chi}\approx\zeta_{\chi}\label{scalar-eigenvectors}
\end{eqnarray}
where $h$ is identified with the observed 125 GeV Higgs boson. The
rotation matrices are defined according to

\begin{eqnarray}
R_{\beta,\alpha} & = & \begin{pmatrix}C_{\beta,\alpha} & S_{\beta,\alpha}\\
-S_{\beta,\alpha} & C_{\beta,\alpha}
\end{pmatrix}.
\end{eqnarray}
The rotation angles are $\beta$, defined as $\tan\beta=T_{\beta}=\frac{\upsilon_{2}}{\upsilon_{1}}$,
and $\alpha$ obtained from the elements of the real mass matrix 
\cite{modelo2} 
\begin{eqnarray}
\tan{2\alpha} & \approx & 
\tan{2\beta}\left[1+2\sqrt{2}S_{\beta}C_{\beta}\left(\frac{\lambda_{2}T_{\beta}^{2}-\lambda_{1}}{T_{\beta}^{2}-1}\right)\left(\frac{\upsilon^{2}}{f_{2}\upsilon_{\chi}}\right)\right]^{-1},\label{alpha-angle}
\end{eqnarray}
where we have taken the dominant contribution assuming that $\upsilon^{2}\ll\left|f_{2}\upsilon_{\chi}\right|$.
In Eq. (\ref{alpha-angle}), we can use the approximation

\begin{equation}
\tan2\alpha\approx\tan2\beta\label{eq:alpha beta}
\end{equation}

\noindent as dominant contribution. After diagonalization of the real
mass matrix, the mass of $\xi_{\chi}$ at dominant order is \cite{modelo3,modelo2}

\begin{equation}
m_{\xi_{\chi}}^{2}\approx2\lambda_{3}\upsilon_{\chi}^{2}.\label{chi mass}
\end{equation}

On the other hand, we obtain all the couplings of the scalar $\chi$
with the above mass eigenstates. The sector of the potential associated
to $\chi$ is:

\begin{eqnarray}
V_{\chi} & = & \mu_{3}^{2}\left|\chi\right|^{2}+\lambda_{3}\left|\chi\right|^{4}+\lambda{}_{6}\left|\chi\right|^{2}\left|\phi_{1}\right|^{2}+\lambda{}_{7}\left|\chi\right|^{2}\left|\phi_{2}\right|^{2}\nonumber \\
 & + & \lambda_{8}\left|\chi\right|^{2}\left|\sigma\right|^{2}.
\end{eqnarray}
After rotation to mass eigenvectors according to (\ref{scalar-eigenvectors}),
we obtain all the interactions of $\chi$ with the scalar matter.
In particular, for the real component $\xi_{\chi}$ of $\chi$, we
obtain:

\begin{eqnarray}
V_{\xi_{\chi}} & = & \frac{1}{2}m_{\xi_{\chi}}^{2}\xi_{\chi}^{2}+\upsilon_{\chi}\xi_{\chi}\left\{ \left(\lambda{}_{6}S_{\beta}^{2}+\lambda{}_{7}C_{\beta}^{2}\right)\left|H^{+}\right|^{2}+\lambda_{8}\xi_{\chi}^{2}|\sigma|^{2}\right.\nonumber \\
 & + & \frac{1}{2}\left(\lambda{}_{6}S_{\alpha}^{2}+\lambda{}_{7}C_{\alpha}^{2}\right)H^{2}+\frac{1}{2}\left(\lambda{}_{6}C_{\alpha}^{2}+\lambda{}_{7}S_{\alpha}^{2}\right)h^{2}\nonumber \\
 & + & \left.\frac{1}{2}\left(\lambda{}_{6}S_{\beta}^{2}+\lambda{}_{7}C_{\beta}^{2}\right)A^{2}\right\} \label{chi-couplings}
\end{eqnarray}

\subsection{Yukawa Lagrangian}

The most general, renormalizable, and $G_{sm}\times U(1)_{X}$ invariant
Yukawa Lagrangian for quarks and with the global symmetry from Eq.
(\ref{global-symm}) is:

\begin{eqnarray}
-\mathcal{L}_{Q} & = & \overline{q_{L}^{1}}\left(\widetilde{\phi}_{2}h_{2}^{U}\right)_{1j}U_{R}^{j}+\overline{q_{L}^{a}}(\widetilde{\phi}_{1}h_{1}^{U})_{aj}U_{R}^{j}\nonumber \\
 & + & \overline{q_{L}^{1}}\left(\phi_{1}h_{1}^{D}\right)_{1j}D_{R}^{j}+\overline{q_{L}^{a}}\left(\phi_{2}h_{2}^{D}\right)_{aj}D_{R}^{j}\nonumber \\
 & + & \overline{q_{L}^{1}}(\phi_{1}h_{1}^{J})_{1m}J_{R}^{m}+\overline{q_{L}^{a}}\left(\phi_{2}h_{2}^{J}\right)_{am}J_{R}^{m}\nonumber \\
 & + & \overline{q_{L}^{1}}\left(\widetilde{\phi}_{2}h_{2}^{T}\right)_{1}T_{R}+\overline{q_{L}^{a}}(\widetilde{\phi}_{1}h_{1}^{T})_{a}T_{R}\nonumber \\
 & + & \overline{T_{L}}\left(\chi^{*}h_{\chi}^{U}\right)_{j}{U}_{R}^{j}+\overline{T_{L}}\left(\chi^{*}h_{T}\right){T}_{R}\nonumber \\
 & + & \overline{J_{L}^{n}}\left(\chi h_{\chi}^{D}\right)_{nj}{D}_{R}^{j}+\overline{J_{L}^{n}}\left(\chi h_{J}\right)_{nm}{J}_{R}^{m}+h.c,\label{quark-yukawa-1}
\end{eqnarray}
where $\widetilde{\phi}_{1,2}=i\sigma_{2}\phi_{1,2}^{*}$ are conjugate
scalar doublets, and $a=2,3$. We can see in Eq. (\ref{quark-yukawa-1})
that due to the non-universality of the $U(1)_{X}$ symmetry, not
all couplings between quarks and scalars are allowed by the gauge
symmetry.

\begin{figure}[t]
\begin{centering}
\includegraphics[scale=0.2]{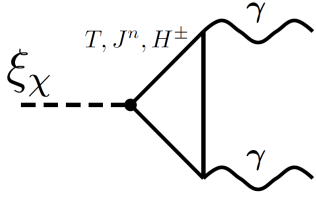}\vspace{-0.5cm}
 
\par\end{centering}

\caption{{\small{}{}Diphoton scalar decay mediated by quarks $T$ ,$J^{n}$
and charged Higgs bosons $H^{\pm}$.}}
\label{fig1} 
\end{figure}

\section{Diphoton decay}

We take the real component $\xi_{\chi}$ of the field $\chi$ as our
750 GeV signal candidate, corresponding to the residual physical particle
after the $U(1)_{X}$ symmetry breaking, while the imaginary component
$\zeta_{\chi}$ corresponds to the would-be Goldstone boson that become
into the longitudinal component of the $Z'$ gauge boson. From the
second term in expression (\ref{chi-couplings}), we can see that
$H^{\pm}$ couples to $\xi_{\chi}$, contributing to the diphoton
decay at one loop level. On the other hand, from the Yukawa Lagrangian
in Eq. (\ref{quark-yukawa-1}) we are interested in the coupling of
the scalar singlet $\chi$, which exhibits couplings with the heavy
sector of the model and mixing terms with the ordinary SM quarks.
The diphoton $\xi_{\chi}$ decay mediated by the $T$, $J^{n}$ quarks
and the charged Higgs boson $H^{\pm}$ is as shown in Fig. \ref{fig1}.

\begin{center}
\begin{figure}[t]
\begin{centering}
\subfloat[]{\includegraphics[scale=0.8]{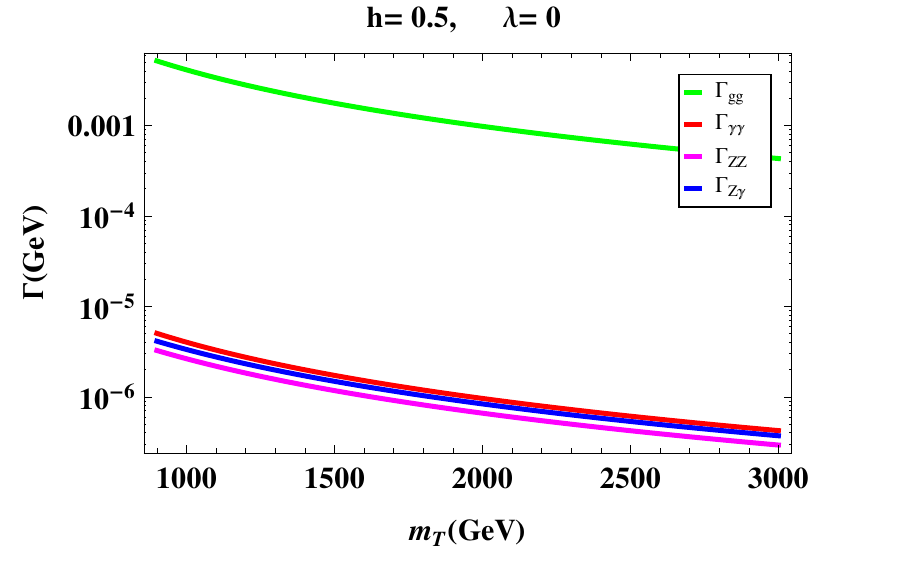} 

}$\qquad$\subfloat[]{\includegraphics[scale=0.8]{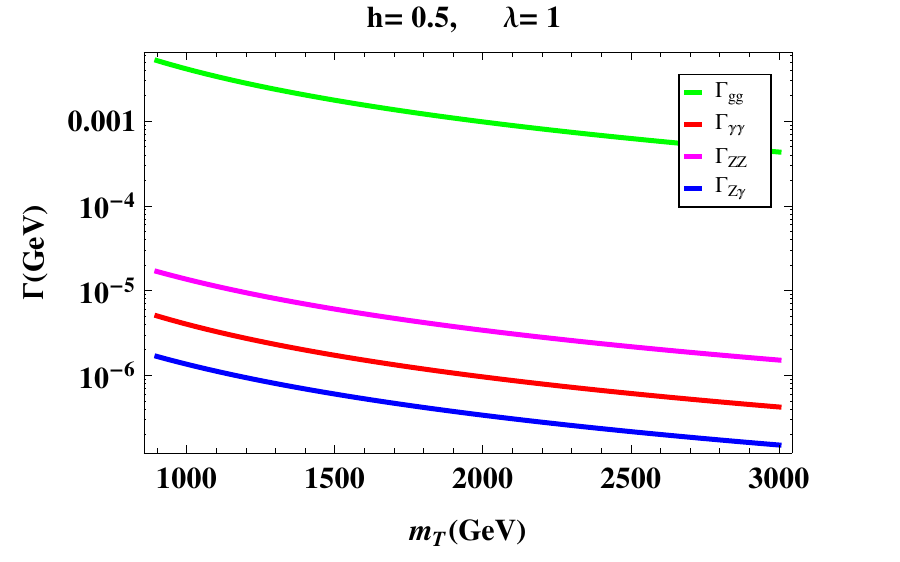} 

}
\par\end{centering}

\caption{Different decay channels for the 750 GeV candidate. \label{fig:decays}}
\end{figure}

\par\end{center}

\subsection{Decay width}

The masses of the extra neutral, pseudoscalar and charged Higgs bosons
$H$, $A$ and $H^{\pm}$, respectively, are nearly degenerated at
the TeV scale, as shown in \cite{modelo3,modelo2}. Then, the decay
of $\xi_{\chi}$ into these Higgs bosons are kinematically forbidden.
The decay into the observed Higgs boson $\xi_{\chi}\rightarrow hh$
is strongly constrained by ATLAS and CMS at 95\%CL \cite{ellis}.
In this way, we obtain the following total decay width for $\xi_{\chi}$:
\begin{eqnarray}
\Gamma & = & \Gamma_{\gamma\gamma}+\Gamma_{gg}+\Gamma_{Z\gamma}+\Gamma_{ZZ}+\Gamma_{\sigma^{*}\sigma}
\end{eqnarray}

We assume the following three scenarios for the total decay width
of our 750 GeV candidate:
\begin{itemize}
\item First, we use one loop contributions, $\Gamma=\Gamma_{\gamma\gamma}+\Gamma_{gg}+\Gamma_{Z\gamma}+\Gamma_{ZZ}$. 
\item Second, we use the experimentally reported width from the ATLAS Collaboration
$\Gamma=45$GeV. 
\item Finally, we consider that the width is dominated by decays into the
scalar dark matter particle $\sigma$, $\Gamma\simeq\Gamma(\xi_{\chi}\to\sigma^{*}\sigma).$ 
\end{itemize}
For the last scenario, after replacing $\upsilon_{\chi}$ in terms
of $m_{\xi_{\chi}}$ from Eq. (\ref{chi mass}), the total decay width
is:

\begin{equation}
\Gamma\simeq\frac{\lambda_{8}m_{\xi_{\chi}}}{32\pi}\sqrt{1-\frac{4m_{\sigma}^{2}}{m_{\xi_{\chi}}^{2}}}.\label{decay-DM}
\end{equation}

For the decay of the $\xi_{\chi}$ particle into one loop contributions,
we consider general interactions of the form

\begin{align}
g_{ZH^{+}H^{-}}=\lambda^{\prime}\left(p_{1}-p_{2}\right)^{\mu},\quad & g_{\gamma H^{+}H^{-}}=\lambda\left(p_{1}-p_{2}\right)^{\mu}.\label{eq:gammacouplings}
\end{align}

We also write the widths in terms of the Yukawa couplings of the top-like
quark $h_{T}$ and bottom-like quarks $h_{J^{1}},\,h_{J^{2}}$, and
the trilinear effective coupling with charged Higgs bosons defined
as

\begin{eqnarray}
h_{H^{\pm}}\equiv\left(\lambda{}_{6}S_{\beta}^{2}+\lambda{}_{7}C_{\beta}^{2}\right),\label{3-chargedhiggs}
\end{eqnarray}

obtaining \cite{Gunion,Cao}:

\begin{align}
\Gamma_{\gamma\gamma} & =\frac{\alpha^{2}m_{\xi_{\chi}}}{32\pi^{3}}\big|\sum_{i}h_{i}N_{i}Q_{i}^{2}F_{i}\big|^{2},\nonumber \\
\Gamma_{gg} & =\frac{\alpha_{s}^{2}m_{\xi_{\chi}}}{16\pi^{3}}\big|\sum_{i\neq H^{\pm}}h_{i}F_{i}\big|^{2},\nonumber \\
\Gamma_{Z\gamma} & =\frac{\alpha^{2}m_{\xi_{\chi}}}{16\pi^{3}}\left(1-\dfrac{m_{Z}^{2}}{m_{\xi_{\chi}}^{2}}\right)^{3}\left|\frac{2}{3}\sum_{i}h_{i}N_{ci}Q_{i}^{2}+\frac{\lambda\lambda^{\prime}}{24\pi\alpha}\right|^{2},\nonumber \\
\Gamma_{ZZ} & =\frac{\alpha^{2}m_{\xi_{\chi}}}{8\pi^{3}}\mathcal{P}\left(\dfrac{m_{Z}^{2}}{m_{\xi_{\chi}}^{2}}\right)\left|\frac{2}{3}\sum_{i}h_{i}N_{ci}Q_{i}^{2}+\frac{\lambda\lambda^{\prime}}{24\pi\alpha}\right|^{2}\label{eq:decays}
\end{align}

\noindent where $\mathcal{P}(x)=\sqrt{1-4x}\left(1-4x+6x^{2}\right)$
is a factor correcting the massive final states in the decay width
and

\[
F_{i}=\begin{cases}
-\sqrt{\tau_{i}}[1+(1-\tau_{i})f(\tau_{i})],\: & i=1/2,\\
\sqrt{\tau_{i}}[1-\tau_{i}f(\tau_{i})] & i=0,
\end{cases}
\]

\noindent with $h_{i}=h_{T},\:h_{J},\:h_{H^{\pm}}$ and $\tau_{i}=4m_{i}^{2}/m_{\xi_{\chi}}^{2}$
for $\tau_{i}>1$, which requires that $m_{i}>375\,\mathrm{GeV}$
for a scalar particle of $m_{\xi_{\chi}}=750\,$GeV. The loop factor
is:

\begin{equation}
f(\tau_{i})=\left[\arcsin\left(\frac{1}{\sqrt{\tau_{i}}}\right)\right]^{2}.\label{loop-factor}
\end{equation}

We emphasize that although the $\xi_{\chi}hh$ coupling is strongly
constrained by ATLAS and CMS data, it does not imply necessarily a
suppression of the $\xi_{\chi}H^{+}H^{-}$ coupling. For example,
if we set:

\begin{align}
\lambda{}_{6}S_{\alpha}^{2}+\lambda{}_{7}C_{\alpha}^{2}\approx & 0,\\
\lambda{}_{6}S_{\beta}^{2}+\lambda{}_{7}C_{\beta}^{2}= & \lambda,
\end{align}

\noindent with $\lambda$ the trilinear effective coupling defined
in (\ref{3-chargedhiggs}), we obtain:

\begin{align}
\lambda_{6}=\frac{-\lambda\,S_{\beta}^{2}}{C_{\beta}^{2}-S_{\alpha}^{2}}, & \quad\lambda_{7}=\frac{\lambda\,C_{\beta}^{2}}{C_{\beta}^{2}-S_{\beta}^{2}},
\end{align}

\noindent where we have used the approximation of Eq. (\ref{eq:alpha beta}),
$\alpha\approx\beta$. In this way, $\xi_{\chi}$ decouple from $hh$
but not from $H^{+}H^{-}$.

\begin{center}
\begin{figure}[t]
\begin{centering}
\subfloat[]{\includegraphics[scale=0.8]{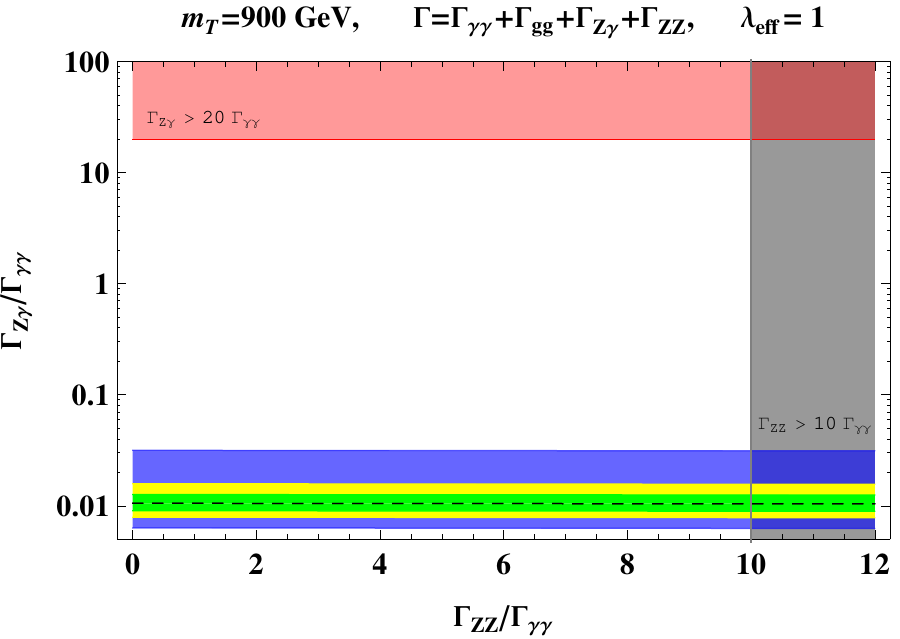} 

}$\qquad$\subfloat[]{\includegraphics[scale=0.8]{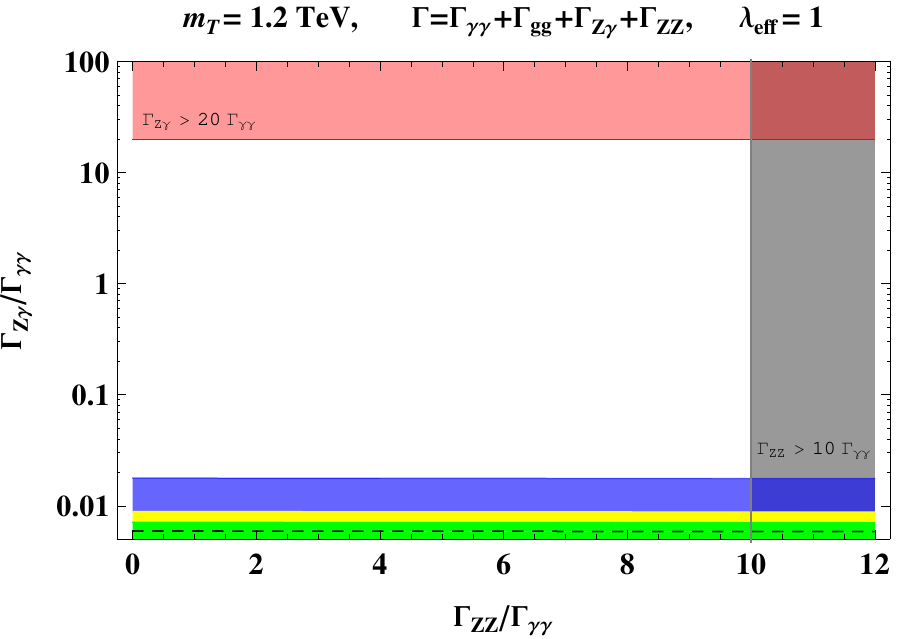} 

}
\par\end{centering}

\begin{centering}
\subfloat[]{\includegraphics[scale=0.8]{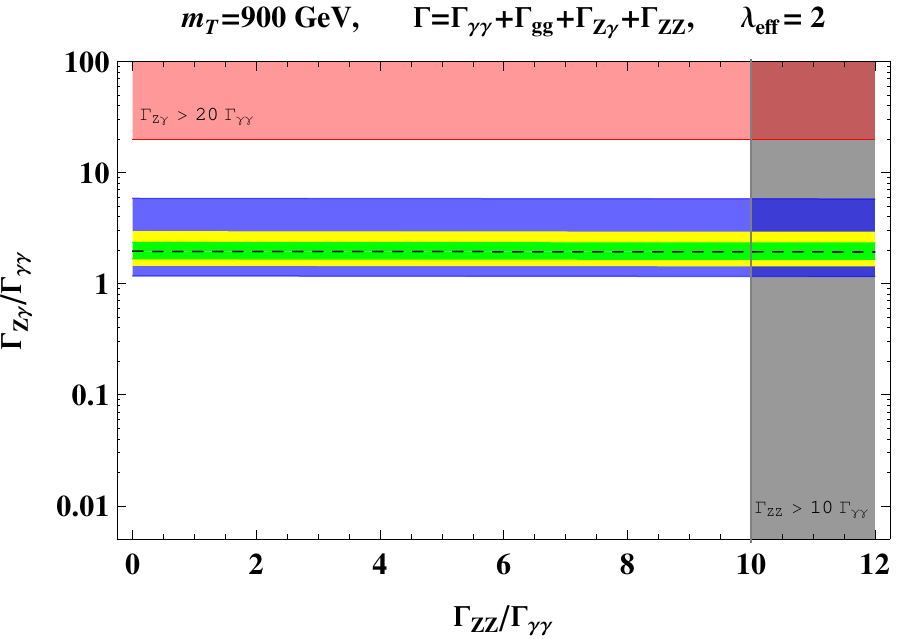} 

}$\qquad$\subfloat[]{\includegraphics[scale=0.8]{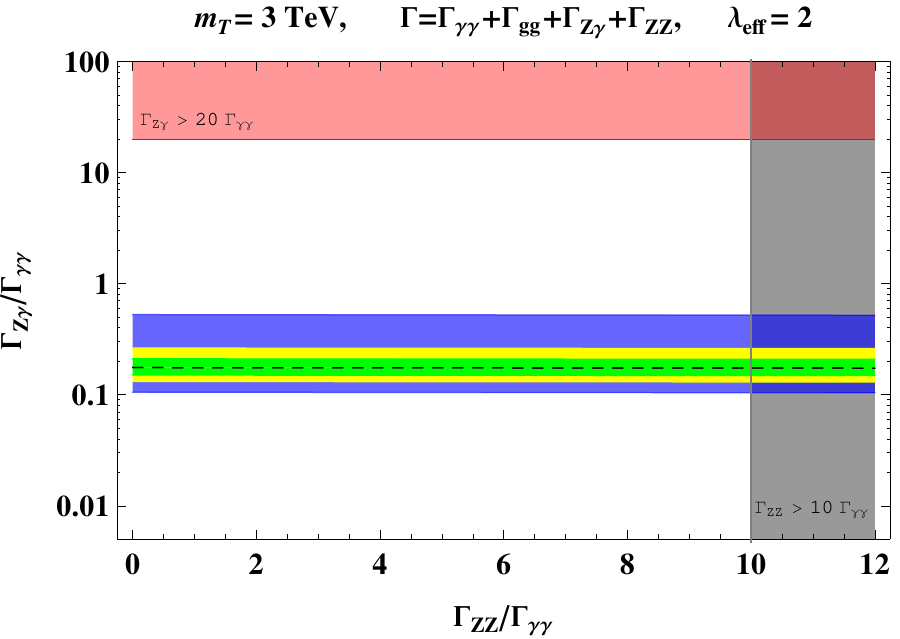} 

}
\par\end{centering}

\caption{Contour plots of the production cross-section $\sigma(pp\to\xi_{\chi}\to\gamma\gamma)$
in femtobarns. The dashed line corresponds to the central value at
6 fb, and the shaded bands corresponds to regions at 68.3\% (green),
95.5\% (yellow) and 99.7\% (light blue) C.L. exclusion limits from
ATLAS and CMS combined data. The shaded red and gray regions are excluded.
\label{fig:exclusionplot} }
\end{figure}

\par\end{center}

\subsection{Production cross section}

\noindent The total cross section $\sigma(pp\to\xi_{\chi}\to\gamma\gamma)$
for a spin zero $\xi_{\chi}$ scalar is given by

\begin{equation}
\sigma(pp\to\xi_{\chi}\to\gamma\gamma)=\frac{C_{gg}\Gamma(\xi_{\chi}\to gg)}{s\:m_{\xi_{\chi}}\Gamma}\Gamma(\xi_{\chi}\to\gamma\gamma),
\end{equation}
where

\begin{equation}
C_{gg}=\frac{\pi^{2}}{8}\intop_{m_{\xi_{\chi}/s}}^{1}\frac{dx}{x}g(x)g(m_{\xi_{\chi}}^{2}/sx)
\end{equation}

\noindent is the dimensionless partonic integral. At the scale $\mu=m_{\xi_{\chi}}=750$
GeV, and center-of-mass energy $\sqrt{s}=13\,\mathrm{TeV}$, this
integral gives $C_{gg}=2137$ \cite{Cgg}. For the analysis, we take
the combined results for the cross section from ATLAS and CMS, $\sigma(pp\to\xi_{\chi}\to\gamma\gamma)=(2-8)\:\mathrm{fb}$
equally valid for $\sqrt{s}=8$ TeV and $C_{gg}=174$ \cite{ellis}.
In addition, we have assumed $\lambda=\lambda^{\prime}$ and $h=h_{T}=h_{J^{1}}=h_{J^{2}}$
for simplicity, i.e, $m_{T}=m_{J^{1}}=m_{J^{2}}$. We have taken $m_{H^{\pm}}=400$
GeV which is the lowest bound reported from charged Higgs boson searches
by ATLAS and CMS \cite{charged higgs}. Also, the lower bound of 900
GeV for $m_{T}$ corresponds to the reported value in recent searches
on top- and bottom-like heavy quarks from ATLAS and CMS Collaborations
\cite{quark masses} and the upper bound of 3 TeV corresponds to the
asymptotic value obtained from the fermionic form factor $F_{1/2}$. 

For the case $\Gamma=\Gamma_{\gamma\gamma}+\Gamma_{gg}+\Gamma_{Z\gamma}+\Gamma_{ZZ}$,
we show in Fig. \ref{fig:decays} the different contributions in Eq.
(\ref{eq:decays}) for the decay width of $\xi$. From Fig. \ref{fig:decays}
(a), the case $\lambda=0$ and $h=0.5$ corresponds to pure fermionic
contributions into the loops. We can see that the contributions (ignoring
the dominant $\Gamma_{gg}$) $\Gamma_{\gamma\gamma},$ $\Gamma_{Z\gamma}$,
$\Gamma_{ZZ}$ have branching ratios of order $42\%,$ $33\%$, $25\%$
respectively. On the other hand, the case $\lambda=0.5$ and $h=0.5$
in Fig. \ref{fig:decays} (b), corresponds to both fermionic and bosonic
contributions into the loop with $\mathrm{BR}_{\gamma\gamma},$ $\mathrm{BR}_{Z\gamma}$,
$\mathrm{BR}_{ZZ}$ of order $22\%,$ $7\%$, $71\%$ respectively. 

In this way, and taking into account current bounds on $\Gamma_{Z\gamma}/\Gamma_{\gamma\gamma}$
and $\Gamma_{ZZ}/\Gamma_{\gamma\gamma}$ \cite{Strumia}, we display
in Fig.\ref{fig:exclusionplot} contour plots of the production cross-section
$\sigma(pp\to\xi_{\chi}\to\gamma\gamma)$ in the $\Gamma_{Z\gamma}/\Gamma_{\gamma\gamma}$-$\Gamma_{ZZ}/\Gamma_{\gamma\gamma}$
plane. For simplicity, we have set $\lambda_{eff}\equiv\lambda=h$
in such a way that the contour plots only depend on $m_{T}$ and $\lambda_{eff}$.
In general, for low values of $m_{T}$ the ratio $\Gamma_{Z\gamma}/\Gamma_{\gamma\gamma}$
is of order $\Gamma_{Z\gamma}/\Gamma_{\gamma\gamma}\sim1$, and for
greater values of $m_{T}$ we have $\Gamma_{Z\gamma}/\Gamma_{\gamma\gamma}<1$.
We also observe that the greater the ratio $\Gamma_{Z\gamma}/\Gamma_{\gamma\gamma}$,
the stronger the coupling $\lambda_{eff}$. However, if $\lambda_{eff}>5$
the model is completely excluded by the bound $\Gamma_{Z\gamma}<20\Gamma_{\gamma\gamma}$
for all $m_{T}$.

\begin{figure}[t]
\begin{centering}
\includegraphics[scale=0.25]{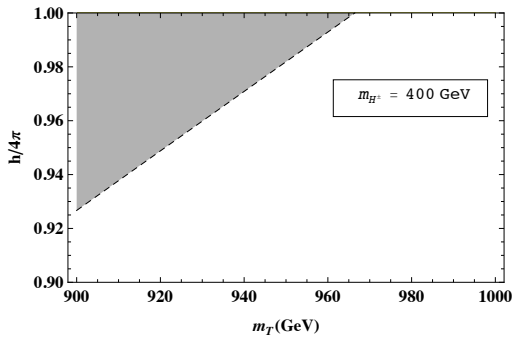} 
\par\end{centering}

\caption{Contours of the production cross-section $\sigma(pp\to\xi_{\chi}\to\gamma\gamma)$
in femtobarns for $\Gamma=45$ GeV.\label{fig:45} The gray region
corresponds to the $99\%$ CL limit.}
\end{figure}

On the other hand, in Fig. \ref{fig:45} we use the width of $\Gamma=45$
GeV reported by the ATLAS Collaboration for the scalar particle of
750 GeV, where we have used $m_{H^{\pm}}=400$ GeV. In this case,
the model is excluded for $m_{T}\geq965$ GeV in the upper limit $h/4\pi=1$
at 99\% CL.

\begin{figure}[t]
\begin{centering}
\subfloat[]{\includegraphics[scale=0.25]{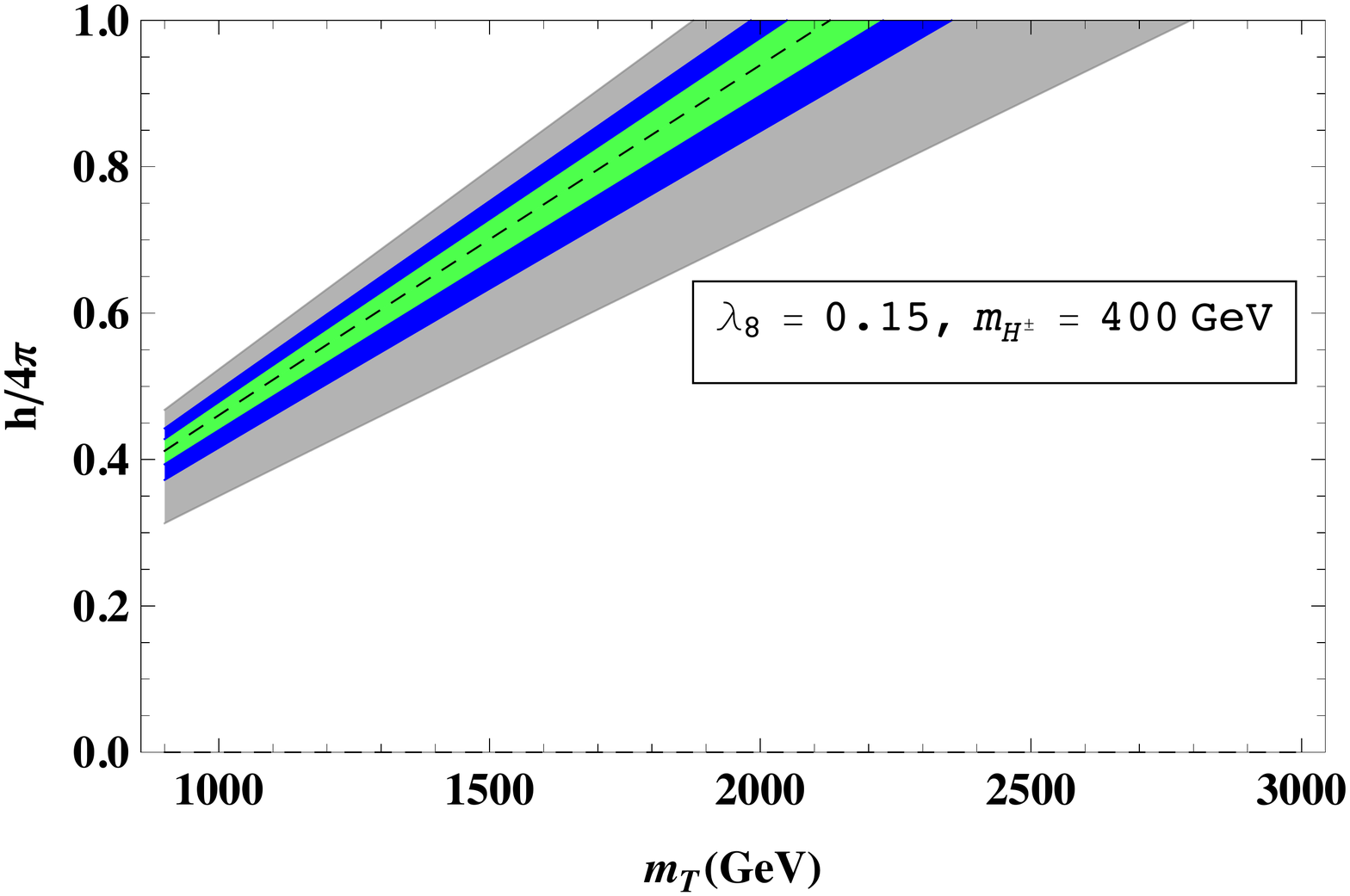} 

}\subfloat[]{\includegraphics[scale=0.25]{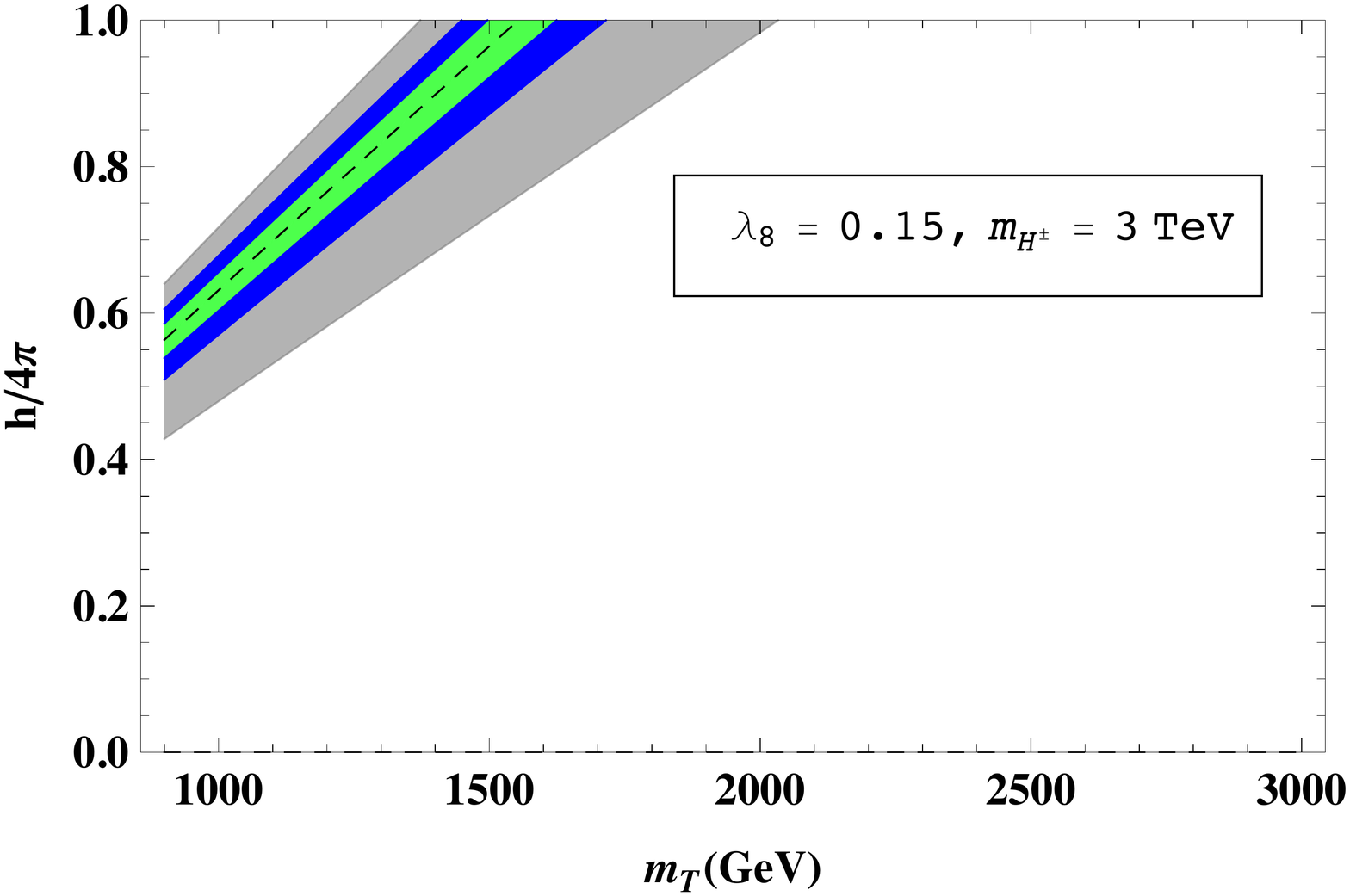} 

}
\par\end{centering}

\centering{}\subfloat[]{\includegraphics[scale=0.25]{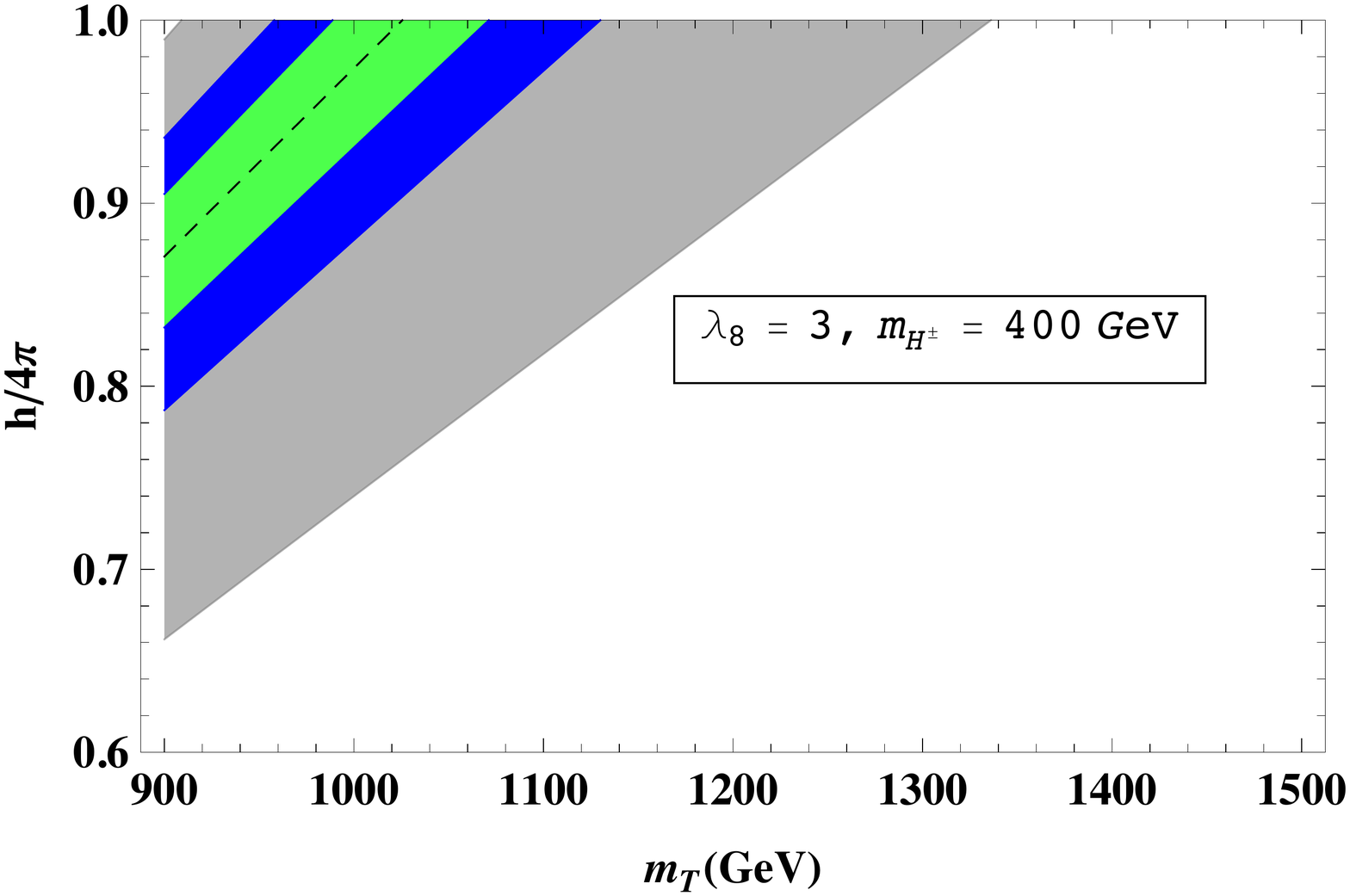} 

}\subfloat[]{\includegraphics[scale=0.25]{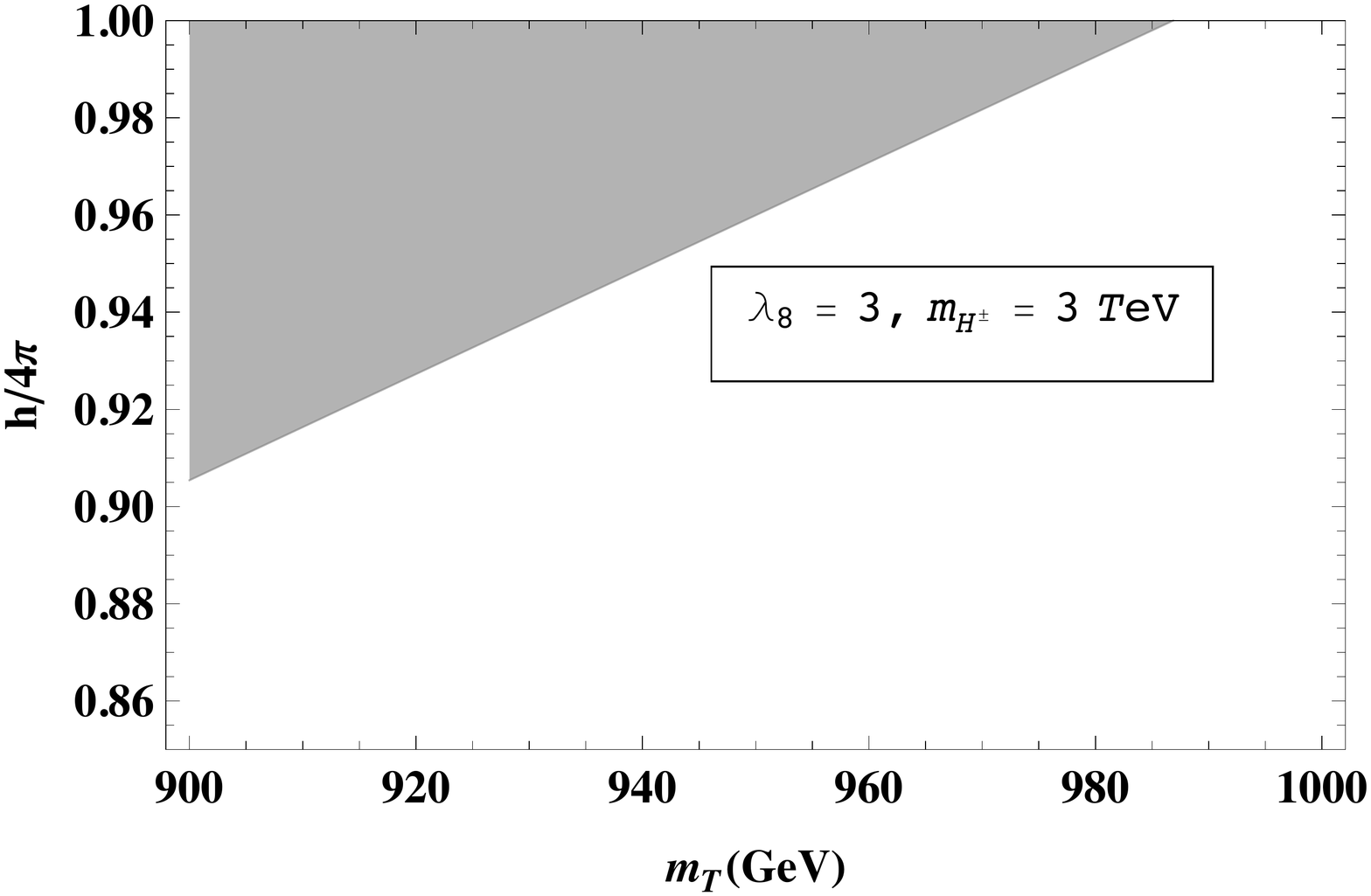}

} \caption{Contour plots of the production cross-section $\sigma(pp\to\xi_{\chi}\to\gamma\gamma)$
in femtobarns with $\Gamma\simeq\Gamma(\xi_{\chi}\to\sigma^{*}\sigma)$.
The shaded gray regions correspond to 99\% CL exclusion limits from
ATLAS and CMS combined data, while the green and blue bands represent
68\% CL and 95\% CL ranges, respectively, around the best fit cross-section
at 6 femtobarns. \label{fig:csDM}}
\end{figure}

Finally, we consider the tree level decay width into the dark matter
candidate of the model, given by Eq. (\ref{decay-DM}). We consider
for the coupling constant, values in the range $0.15\leq\lambda_{8}\leq3$
and for the decay width in the range $1.2\,\textrm{GeV}\leq\Gamma\leq23\,\textrm{GeV}$.
If $\lambda_{8}\sim0.15$, the width of the $750$ GeV candidate is
$\Gamma(\xi_{\chi}\to\sigma^{*}\sigma)\sim1$GeV. Thus, this decay
channel become in the dominant contribution, larger than the loop
contributions. We also see that the dark matter decay width is sensitive
to its mass $m_{\sigma}$ only near to the kinematical region. In
Fig. \ref{fig:csDM}, we show the production cross section contours
GeV. In Figs. \ref{fig:csDM}(a) and (b), we set $\lambda_{8}=0.15$
and a decay width of $\Gamma=1.2$ GeV for $m_{H^{\pm}}=400$ GeV
and $m_{H^{\pm}}=3.0$ TeV respectively. For this set of parameters
in Figs. \ref{fig:csDM}(a) and (b) the model is excluded for exotic
quark masses greater than 2.7 TeV and 2.0 TeV respectively. In Figs.
\ref{fig:csDM}(c) and (d) we set $\lambda_{8}=3$ with a decay width
of $\Gamma=23$ GeV and the same values for $m_{H^{\pm}}$ as before.
In this case the model is excluded for $m_{T}>1.3$ TeV in Fig. \ref{fig:csDM}(c)
and $m_{T}>0.98$ TeV in Fig. \ref{fig:csDM}(d).

\section{Conclusions}

Since the announcement of the ATLAS and CMS collaborations of a possible
750 GeV diphoton excess, many authors have attempted to explain the
signal in the framework of several extensions of the SM that includes
some type of resonance compatible with the reported data. In this
work, we use a well-founded nonuniversal $U(1)_{X}$ extension that
includes an extra particle sector, with a neutral scalar singlet being
the candidate for the 750 GeV signal. Finding non-trivial solutions
for the $U(1)_{X}$ charge that cancel the chiral anomalies, the model
requires an structure of three fermion families, and an extension
of the quark sector, being the most simple one top-like and two bottom-like
quasi chiral singlets. In addition, the model contains two Higgs doublets
in order to provide masses to all fermions. In particular, after the
symmetry breaking, one charged Higgs boson that couple with the scalar
singlet is obtained. Thus, in a natural way, the model predicts a
diphoton decay of the scalar singlet through one-loop corrections
mediated by quark singlets and a charged Higgs boson. Finally, we
include another scalar singlet with a $U(1)$ global symmetry as candidate
for dark matter, and that also may couple with the 750 GeV scalar
at tree level, contributing to the decay width. We found allowed regions
in different scenarios compatible with a $750$ GeV signal for masses
of the top-like quark in the range $0.9<m_{T}<3$ TeV and charged
Higgs bosons at $0.4$ and $3$ TeV.

\section*{Acknowledgment}

This work was supported by El Patrimonio Autónomo Fondo Nacional de
Financiamiento para la Ciencia, la Tecnología y la Innovación Francisco
José de Caldas programme of COLCIENCIAS in Colombia.

\end{document}